\documentclass{ws-ijmpa}

\usepackage[super]{cite}
\usepackage{xcolor}
\usepackage[verbose,hypertexnames=false]{hyperref}
\hypersetup{colorlinks=false,allbordercolors=blue,pdfborderstyle={/S/U/W 1}}

\begin{document}

\markboth{M. Szilágyi, M. Csanád, and D. Kincses}{Methods for Extracting 3D Lévy Source Parameters}

%
\catchline{}{}{}{}{}
%

\title{Comparison of Methods for Extracting Three-Dimensional Lévy Source Parameters in $\sqrt{s_{\text{NN}}} = 200\text{ GeV}$ Au+Au Collisions with EPOS3}

\author{Máté Szilágyi, Máté Csanád and Dániel Kincses}

\address{Department of Atomic Physics, ELTE Eötvös Loránd University,\\
Pázmány Péter sétány 1/A, Budapest, H-1117, Hungary\\
E-mail: szilagyi03@staff.elte.hu, csanad@elte.hu, kincses@ttk.elte.hu}

\maketitle

\begin{history}
\received{Day Month Year}
\revised{Day Month Year}
\accepted{Day Month Year}
\published{Day Month Year}
\end{history}

\begin{abstract}
Bose--Einstein femtoscopy probes the spatiotemporal geometry of the particle-emitting source in heavy-ion collisions, where experimental data favor a Lévy-stable source description. We compare four approaches for extracting 3D Lévy source parameters using $\pi^+\pi^+$ pairs from EPOS3-simulated Au+Au collisions at $\sqrt{s_{\text{NN}}} = 200\text{ GeV}$: direct fits to the pair-source distribution $D(\bm{\rho})$, and fits to the two-particle correlation function $C(\bm{q})$ obtained via (i) the CRAB package, (ii) direct numerical integration of $D(\bm{\rho})$, and (iii) fast Fourier transformation (FFT). On an identical momentum grid, numerical integration and FFT are analytically identical and numerically equivalent ($\text{RMSE} < 10^{-14}$), while FFT reduces the computation time from hours to seconds.

We extract the transverse-mass dependence of the Lévy radii, Lévy index $\alpha$, and correlation strength $\lambda$ in $0\text{--}10\,\%$ central collisions. All methods reproduce the same physical trends, but CRAB results fluctuate due to event pooling, and correlation-function fits deviate systematically from direct source fits, e.g., with $\sim 7\text{--}9\,\%$ lower $\alpha$. These deviations may reflect residual non-Lévy features of the source, probed differently in coordinate and momentum space, and the comparison of event-averaged parameters with fits to the ensemble-accumulated source, a mixture of Lévy distributions with lower effective $\alpha$.
\end{abstract}


\ccode{PACS numbers: 25.75.-q, 25.75.Gz, 05.40.Fb}

\section{Introduction}

Ultra-relativistic heavy-ion collisions create an extreme state of matter known as the quark--gluon plasma (QGP), which undergoes rapid hydrodynamic expansion and cooling until quarks and gluons hadronize. Following chemical freeze-out, where inelastic processes cease, the produced hadrons continue to interact elastically and resonances decay until a kinetic freeze-out or complete kinetic decoupling, after which the given particle streams freely toward the detectors. Because the spatial and temporal extent of the emitting medium lies on the femtometer scale ($1\text{ fm} = 10^{-15}\text{ m}$), its spacetime geometry cannot be resolved by direct imaging. Instead, Bose--Einstein femtoscopy exploits quantum-statistical momentum correlations of identical bosons to map the spatiotemporal structure of the particle-emitting source at kinetic freeze-out.

The two-particle momentum correlation function is defined as the ratio of the two-particle momentum distribution $N_2(p_1, p_2)$ to the product of the single-particle momentum distributions $N_1(p_1)$ and $N_1(p_2)$:
\begin{equation}
C_2(p_1, p_2) = \frac{N_2(p_1, p_2)}{N_1(p_1) N_1(p_2)}\,,
\label{eq:C2_def}
\end{equation}
where $p_1$ and $p_2$ denote the four-momenta of the two particles. To connect the momentum distributions to the spatial geometry of the emission region, one introduces the single-particle phase-space emission function (source function) $S(x, p)$, which represents the probability density of producing a particle with four-momentum $p$ at spacetime four-coordinate $x$. In terms of the single-particle wavefunction $\Psi(x_1)$ and the two-particle wavefunction $\Psi_2(x_1, x_2)$, the inclusive momentum distributions take the form
\begin{align}
N_1(p_1) &= \int S(x_1, p_1) |\Psi(x_1)|^2 \, d^4x_1\,, \label{eq:N1_def} \\
N_2(p_1, p_2) &= \int S(x_1, p_1) S(x_2, p_2) |\Psi_2(x_1, x_2)|^2 \, d^4x_2 \, d^4x_1\,. \label{eq:N2_def}
\end{align}
Neglecting final-state Coulomb and strong interactions as well as multi-particle correlations, the single-particle propagation is described by a plane wave, $\Psi(x) = e^{ipx}$, with $|\Psi(x)|^2 = 1$, where we used natural units ($\hbar = 1, c = 1$). For two indistinguishable bosons, Bose--Einstein quantum statistics requires a symmetrized two-particle wavefunction,
\begin{equation}
\Psi_2(x_1, x_2) = \frac{1}{\sqrt{2}} \left( e^{i(p_1 x_1 + p_2 x_2)} + e^{i(p_1 x_2 + p_2 x_1)} \right)\,,
\label{eq:psi2}
\end{equation}
whose squared modulus depends only on the relative four-momentum $q = p_1 - p_2$ and the relative four-coordinate $x_1 - x_2$:
\begin{equation}
|\Psi_2(x_1, x_2)|^2 = 1 + \cos\big((p_1 - p_2)(x_1 - x_2)\big) = 1 + \cos\big(q(x_1 - x_2)\big)\,.
\label{eq:psi2_sq}
\end{equation}

Introducing the pair average four-momentum $K = (p_1 + p_2)/2$ alongside $q = p_1 - p_2$, one notes that quantum-statistical correlations are predominantly concentrated in the kinematic region where $K \gg q$. This justifies the smoothness approximation $p_1 \approx p_2 \approx K$ in the arguments of the source functions\cite{Lisa2005}. Defining the Fourier transform of the source function as $\tilde{S}(q, K) = \int S(x, K) e^{iqx} \, d^4x$ and noting that $\int S(x, K) \, d^4x = \tilde{S}(q = 0, K)$, insertion of Eq.~(\ref{eq:psi2_sq}) into Eq.~(\ref{eq:N2_def}) yields $N_2^{(0)}(q, K) = |\tilde{S}(q = 0, K)|^2 + |\tilde{S}(q, K)|^2$, while $N_1(p_1) N_1(p_2) \approx |\tilde{S}(q = 0, K)|^2$. Consequently, the interaction-free correlation function (indicated by the superscript $(0)$) becomes
\begin{equation}
C_2^{(0)}(q, K) = 1 + \frac{|\tilde{S}(q, K)|^2}{|\tilde{S}(q = 0, K)|^2}\,.
\label{eq:C2_S}
\end{equation}
This relation can be expressed directly in terms of the pair-source distribution $D(\rho, K)$ (also denoted as $D(x, K)$ in four-vector notation), which represents the autocorrelation of the single-particle emission function:
\begin{equation}
D(\rho, K) = \int S\left(X + \frac{\rho}{2}, K\right) S\left(X - \frac{\rho}{2}, K\right) d^4X\,,
\label{eq:D_def}
\end{equation}
where $X = (x_1 + x_2)/2$ is the pair center-of-mass four-coordinate and $\rho = x_1 - x_2$ is the four-dimensional relative spacetime separation of the pair. Denoting the Fourier transform of the pair-source distribution by $\tilde{D}(q, K)$, Eq.~(\ref{eq:C2_S}) simplifies to
\begin{equation}
C_2^{(0)}(q, K) = 1 + \frac{\tilde{D}(q, K)}{\tilde{D}(q = 0, K)}\,,
\label{eq:C2_D}
\end{equation}
which establishes the fundamental Fourier link between the spatial pair-source distribution and the two-particle momentum correlation function.

For identical particles of mass $m$, the four-vectors $q$ and $K$ are orthogonal, since $q \cdot K = (p_1 - p_2)(p_1 + p_2)/2 = (p_1^2 - p_2^2)/2 = (m^2 - m^2)/2 = 0$. Writing $q \cdot K = q_0 K_0 - \bm{q} \cdot \bm{K} = 0$ allows the energy difference $q_0$ to be eliminated via the three-vectors $\bm{q} \equiv (q_x, q_y, q_z)$ and $\bm{K} \equiv (K_x, K_y, K_z)$ as
\begin{equation}
q_0 = \frac{\bm{q} \cdot \bm{K}}{K_0}\,.
\label{eq:q0_elim}
\end{equation}
The correlation function is therefore fully parameterized by the three-momenta as $C_2(\bm{q}, \bm{K})$. Because the dependence of the correlation function\cite{Csorgo2002} on $\bm{K}$ is significantly smoother than that on $\bm{q}$, measurements are performed by dividing the pair sample into narrow intervals of $\bm{K}$ and fitting the $\bm{q}$-dependence within each bin. At midrapidity ($y = \frac{1}{2} \ln \frac{E + p_z}{E - p_z} \approx 0$), the kinematic dependence of the extracted source parameters is investigated as a function of the average transverse momentum $K_{\text{T}}$ or the transverse mass $m_{\text{T}}$, defined as
\begin{equation}
K_{\text{T}} = \sqrt{K_x^2 + K_y^2}\,, \qquad m_{\text{T}} = \sqrt{m^2 + K_{\text{T}}^2}\,.
\label{eq:KT_mT}
\end{equation}
To separate the spatial and temporal components of the source, the relative momentum $\bm{q}$ is decomposed in the right-handed Bertsch--Pratt coordinate system\cite{Pratt1986,Bertsch1989} into $\bm{q}_{\text{BP}} \equiv (q_{\text{out}}, q_{\text{side}}, q_{\text{long}})$, where the ``long'' axis points along the beam direction, the ``out'' axis is parallel to the pair average transverse momentum, and the ``side'' axis is perpendicular to both. Evaluating the kinematics in the Longitudinal Comoving System (LCMS), where the longitudinal component of the pair average momentum vanishes, one has $\bm{K}_{\text{BP,LCMS}} = (K_{\text{T}}, 0, 0)$. In this frame, Eq.~(\ref{eq:q0_elim}) reduces to $q_0 = q_{\text{out}} K_{\text{T}} / K_0$, demonstrating that the emission duration of the source is coupled exclusively to the ``out'' direction; the corresponding effective three-dimensional reduced pair separation in the LCMS is denoted hereafter by boldface $\bm{\rho}$ (see Sec.~2.4 for its explicit coordinate representation), in distinction to the four-vector $\rho$ in Eq.~(\ref{eq:D_def}).

Although Eqs.~(\ref{eq:C2_S}) and (\ref{eq:C2_D}) imply the exact intercept $C_2^{(0)}(\bm{q} \to \bm{0}, \bm{K}) = 2$, finite two-track resolution in experiments prevents access to relative momenta below $q_{\text{min}} \approx 6\text{--}8\text{ MeV}/c$. Extrapolating the measured correlation function to $\bm{q} \to \bm{0}$ defines the correlation strength (intercept parameter)\cite{Csorgo1996} $\lambda(\bm{K}) \equiv \lim_{\bm{q} \to \bm{0}} C_2(\bm{q}, \bm{K}) - 1$, which typically satisfies $\lambda < 1$. This reduction is naturally explained by the core--halo model,\cite{Csorgo1996} which decomposes the single-particle source into a hydrodynamically expanding primordial ``core'' of characteristic size $10\text{--}20\text{ fm}$ and an extended ``halo'' of decay products from long-lived resonances (such as $\eta$ mesons) reaching sizes well beyond a few hundred $\text{fm}$ (up to $\sim 10^5\text{ fm}$): $S(x, \bm{K}) = S_{\text{core}}(x, \bm{K}) + S_{\text{halo}}(x, \bm{K})$. Because the Fourier transform of the broad halo is confined to the unresolvable region $q < q_{\text{min}}$, one has $\tilde{S}(\bm{q}, \bm{K}) \approx \tilde{S}_{\text{core}}(\bm{q}, \bm{K})$ in the measurable momentum range. Similarly, in the decomposition of the pair source, $D(\bm{\rho}, \bm{K}) = D_{\text{(core,core)}}(\bm{\rho}, \bm{K}) + D_{\text{(core,halo)}}(\bm{\rho}, \bm{K}) + D_{\text{(halo,halo)}}(\bm{\rho}, \bm{K})$, only the core--core term contributes at resolvable $\bm{q}$, leading to
\begin{equation}
C_2^{(0)}(\bm{q}, \bm{K}) \approx 1 + \lambda(\bm{K}) \frac{\tilde{D}_{\text{(core,core)}}(\bm{q}, \bm{K})}{\tilde{D}_{\text{(core,core)}}(\bm{0}, \bm{K})}\,, \qquad \lambda(\bm{K}) = \left( \frac{N_{\text{core}}(\bm{K})}{N_{\text{core}}(\bm{K}) + N_{\text{halo}}(\bm{K})} \right)^2\,,
\label{eq:core_halo}
\end{equation}
where $N_{\text{core}}(\bm{K}) = \tilde{S}_{\text{core}}(\bm{0}, \bm{K})$ and $N_{\text{halo}}(\bm{K}) = \tilde{S}_{\text{halo}}(\bm{0}, \bm{K})$ denote the respective particle multiplicities.

While early femtoscopic analyses assumed a Gaussian source profile, experimental measurements\cite{PHENIX:2024,PHENIX:2017ino,CMS:2023xyd,NA61SHINE:2023qzr,Kincses:2024sin} revealed power-law tails in the correlation function that are accurately captured by generalizing the source to a three-dimensional symmetric L\'evy-stable distribution,\cite{CsorgoHegyiZajc2004,Csanad2024} physically motivated for instance by anomalous diffusion during hadronic rescattering\cite{CsanadCsorgoNagy2007,kincses2022,kincses2025}. The three-dimensional symmetric L\'evy-stable distribution is defined via its inverse Fourier transform as
\begin{equation}
\mathcal{L}(\alpha, R^2, \bm{x}) = \frac{1}{(2\pi)^3} \int d^3\bm{q} \, e^{i\bm{q} \cdot \bm{x}} \exp\left( -\frac{1}{2} |\bm{q}^T R^2 \bm{q}|^{\alpha/2} \right)\,,
\label{eq:Levy_def}
\end{equation}
where $R^2 = \text{diag}(R_{\text{out}}^2, R_{\text{side}}^2, R_{\text{long}}^2)$ is the diagonal matrix of squared L\'evy scale parameters in the Bertsch--Pratt frame, and $\alpha$ is the L\'evy stability index. The case $\alpha = 2$ recovers the Gaussian distribution, whereas $\alpha < 2$ generates power-law spatial tails. Because the autocorrelation of a L\'evy-stable distribution preserves its functional form, a core emission function $S_{\text{core}}(\bm{x}) = \mathcal{L}(\alpha, R^2, \bm{x})$ yields a L\'evy-stable core pair-source distribution with rescaled radii:
\begin{equation}
D_{\text{(core,core)}}(\bm{\rho}) = \mathcal{L}\left(\alpha, 2^{2/\alpha} R^2, \bm{\rho}\right)\,.
\label{eq:D_Levy}
\end{equation}
Its Fourier transform is $\tilde{D}_{\text{(core,core)}}(\bm{q}) = \exp(-\vert{}\bm{q}^T R^2 \bm{q}\vert{}^{\alpha/2})$. Substituting this into Eq.~(\ref{eq:core_halo}) yields the analytical form of the three-dimensional two-particle correlation function for a L\'evy-stable source:
\begin{equation}
C_2^{(0)}(q_{\text{out}}, q_{\text{side}}, q_{\text{long}}) = 1 + \lambda \exp\left( -\left| q_{\text{out}}^2 R_{\text{out}}^2 + q_{\text{side}}^2 R_{\text{side}}^2 + q_{\text{long}}^2 R_{\text{long}}^2 \right|^{\alpha/2} \right)\,,
\label{eq:C2_Levy_3D}
\end{equation}
which contains five physical parameters to be extracted in each $K_{\text{T}}$ interval: the correlation strength $\lambda$, the three L\'evy radii $R_{\text{out}}, R_{\text{side}}, R_{\text{long}}$, and the L\'evy index $\alpha$.

\section{Data and Methodology}

\subsection{Simulated data and kinematic selections}

In this work, we analyze $\text{Au}+\text{Au}$ collisions at a center-of-mass energy of $\sqrt{s_{\text{NN}}} = 200\text{ GeV}$ simulated with the EPOS3 event generator.\cite{Werner2010,kincses2022} EPOS3 incorporates initial-state fluctuations via Parton Gribov--Regge theory, a core--corona division, viscous hydrodynamical expansion, hadronization, hadronic rescattering, and resonance decays. The simulation output provides the complete phase-space emission coordinates $(x, p)$ at kinetic freeze-out, though without quantum-statistical Bose--Einstein symmetrization.

To enable a direct and consistent comparison with the coordinate-space source-fitting studies of Kincses et al.,\cite{kincses2025,kincses2026lambda} we adopt the exact same centrality selection and single-particle acceptance criteria. Specifically, the $0\text{--}10\,\%$ most central events (comprising approximately $30\,000$ simulated EPOS3 events) are selected based on the collision impact parameter ($b$) distribution. The analysis is restricted to positively charged pions ($\pi^+$) satisfying the midrapidity single-particle cuts $|\eta| < 1.0$ (where $\eta = \frac{1}{2}\ln\frac{|\bm{p}|+p_z}{|\bm{p}|-p_z}$) and $0.15\text{ GeV}/c < p_{\text{T}} < 1.0\text{ GeV}/c$. Furthermore, strictly following the binning of Refs.~\refcite{kincses2025} and \refcite{kincses2026lambda}, the pion pairs are divided into eight average transverse-momentum ($K_{\text{T}}$) intervals: $[175\text{--}225, 225\text{--}275, 275\text{--}325, 325\text{--}375, 375\text{--}425, 425\text{--}475, 475\text{--}525, 525\text{--}575]\text{ MeV}/c$, with the corresponding transverse mass $m_{\text{T}}$ evaluated at each bin center via Eq.~(\ref{eq:KT_mT}).

\subsection{Benchmark method: Direct pair-source fitting}

As a primary baseline for our momentum-space femtoscopic investigations, we compare our results with the coordinate-space benchmark analysis established by Kincses, Nagy, and Csan\'ad in Ref.~\refcite{kincses2025} and continued in Ref.~\refcite{kincses2026lambda} (from which our reference dataset and systematic uncertainty bands are taken). In those works, the authors extracted three-dimensional L\'evy source parameters directly from the spatial freeze-out coordinates of the identical EPOS3 dataset, bypassing the construction of momentum correlation functions entirely.

In their approach, the three-dimensional pair-source distribution $D(\bm{\rho}, \bm{K})$ was reconstructed on an event-by-event basis from the relative spatial coordinates $\bm{\rho}$ of all accepted pion pairs (including resonance decay products) in the LCMS frame and normalized to unity over the full spatial domain (extending up to $\sim 10^5\text{ fm}$). Its one-dimensional projections along the Bertsch--Pratt axes were then fitted simultaneously within the $m_{\text{T}}$-dependent spatial region $|\rho_i| < \rho_{\text{fit,max}} = \sqrt{B / m_{\text{T}}}$ with $B = 2500\text{ fm}^2\,\text{GeV}/c^2$ using the corresponding projections of the theoretical three-dimensional L\'evy core distribution multiplied by the intercept parameter, $\lambda\,\mathcal{L}(\alpha, 2^{2/\alpha} R^2, \bm{\rho})$.\cite{Kincses2025shapes,kincses2026lambda} Because pairs involving decay products of long-lived resonances (such as $\eta, \eta'$, and $K^0_S$ mesons) populate large separations well beyond $\rho_{\text{fit,max}}$ and have negligible density inside the fit window while still contributing to the unit normalization of $D(\bm{\rho}, \bm{K})$ over all distances, the fitted normalization prefactor $\lambda$ directly measures the core--core pair fraction (i.e., the integral of the fitted L\'evy core relative to the full pair distribution).\cite{Kincses2025shapes,kincses2026lambda} Averaging the fitted parameters over the event sample then yielded the final event-averaged L\'evy radii $\langle R_{\text{out}} \rangle, \langle R_{\text{side}} \rangle, \langle R_{\text{long}} \rangle$, stability index $\langle \alpha \rangle$, and correlation strength $\langle \lambda \rangle$.

\subsection{Correlation function construction and fitting with CRAB}

To model the experimental extraction of source parameters in momentum space, we employ the Correlation After Burner (CRAB) software package,\cite{Pratt1994} which generates the two-particle correlation function $C(\bm{q})$ from the simulated freeze-out phase-space distribution $S(x, p)$. Following experimental practice, CRAB constructs the correlation function as the ratio
\begin{equation}
C(\bm{q}) = \frac{A(\bm{q})}{B(\bm{q})}\,,
\label{eq:CRAB_AB}
\end{equation}
where $A(\bm{q})$ is the relative-momentum distribution of correlated particle pairs, and $B(\bm{q})$ is a reference background distribution of uncorrelated pairs that preserves single-particle acceptance and phase-space constraints.

In our implementation, CRAB first reads the simulated events and stores all $\pi^+$ candidates satisfying the single-particle cuts ($0.15\text{ GeV}/c < p_{\text{T}} < 1.0\text{ GeV}/c$ and $\vert{}\eta\vert{} < 1.0$). To enhance pair statistics, events with similar impact parameters $b$ are grouped together and merged into pooled event classes with an average impact parameter. Next, for both $A(\bm{q})$ and $B(\bm{q})$ within each $K_{\text{T}}$ bin, the program randomly samples $N = 10^8$ pion pairs from these pooled event classes that satisfy the pair kinematic cuts $K_{\text{min}} < K_{\text{T}} < K_{\text{max}}$ and $0 < q_i < 150\text{ MeV}/c$ along each axis. Consequently, the numerator pairs in $A(\bm{q})$ are also drawn from the merged event classes (and weighted with the quantum-statistical weight) rather than strictly from the same individual EPOS3 collision event, meaning that $A(\bm{q})$ does not sample purely the within-event pair source $D(\bm{\rho})$. Furthermore, because the grouping of events by impact parameter leaves only a small number of distinct event blocks, standard event mixing across different event blocks is replaced when generating $B(\bm{q})$ by rotating both the position and momentum coordinates of one particle in the pair by a random azimuthal angle in the transverse plane.

For each accepted pair, the quantum-statistical weight $\vert{}\Psi_{\text{pair}}\vert{}^2$ is evaluated and accumulated into the corresponding bin of the three-dimensional histograms $A(\bm{q})$ and $B(\bm{q})$. For the numerator $A(\bm{q})$, the weight is calculated according to Eq.~(\ref{eq:psi2_sq}) with explicit unit conversion:
\begin{equation}
|\Psi_{\text{pair}}|^2 = 2 \cos^2\left( \frac{\bm{q} \cdot \bm{r}}{2 \hbar c} \right) = 2 \cos^2\left( \frac{\bm{q} \cdot \bm{r}}{2 \cdot 197.327\text{ MeV fm}} \right)\,,
\label{eq:CRAB_weight}
\end{equation}
whereas for the uncorrelated background $B(\bm{q})$, unit weight $\vert{}\Psi_{\text{pair}}\vert{}^2 = 1$ is assigned. Both distributions are histogrammed on a three-dimensional Cartesian grid in the Bertsch--Pratt LCMS frame with $100$ bins along each axis ($q_{\text{out}}, q_{\text{side}}, q_{\text{long}}$) spanning the range $0 < q_i < 150\text{ MeV}/c$, after which $C(\bm{q})$ and its statistical uncertainties are obtained via bin-by-bin division. The CRAB computation requires several hours of execution time per $K_{\text{T}}$ bin, with the exact duration varying between $K_{\text{T}}$ bins.

Subsequently, the resulting three-dimensional correlation function $C(\bm{q})$ is fitted with the analytical L\'evy parametrization of Eq.~(\ref{eq:C2_Levy_3D}) (incorporating the $\hbar c$ conversion factor in the exponent) within a spherical fit domain $\vert{}\bm{q}\vert{}_{\text{max}} = 200\text{ MeV}/c$ beyond which pair statistics become insufficient. Fitting only the positive octant with the diagonal parameterization of Eq.~(\ref{eq:C2_Levy_3D}) assumes reflection symmetry in each $q_i$ component separately (i.e., a vanishing out--long cross term), which is justified by the symmetric $|\eta| < 1.0$ acceptance window around midrapidity ($y \approx 0$). Because $C(\bm{q}) = A(\bm{q})/B(\bm{q})$ is constructed as the ratio of two Poisson-distributed histograms (with $\approx 100\text{--}200$ pairs per three-dimensional bin), standard least-squares ($\chi^2$) minimization performs poorly. Instead, we determine the best-fit parameters ($\lambda, R_{\text{out}}, R_{\text{side}}, R_{\text{long}}, \alpha$) by minimizing the Poisson log-likelihood statistic\cite{Baker1984,Ahle2002}:
\begin{equation}
\chi^2_{\text{log}} = -2 \sum \left[ A \ln\left( \frac{f(A + B)}{A(1 + f)} \right) + B \ln\left( \frac{A + B}{B(1 + f)} \right) \right]\,,
\label{eq:chi2_log}
\end{equation}
where $A \equiv A(\bm{q})$, $B \equiv B(\bm{q})$, and $f \equiv C_2^{(0)}(\bm{q})$ denotes the fitted correlation function evaluated at the bin center. Finally, the goodness-of-fit (confidence level) is quantified from the standard $\chi^2 = \sum [(C - f)/\sigma]^2$ divided by the number of degrees of freedom (NDF).

\subsection{Numerical integration of the pair-source distribution}

An alternative avenue to extract source parameters from transport simulations without relying on random pair sampling from pooled event classes is the direct numerical integration of the Koonin--Pratt formula.\cite{Koonin1977,PrattCsorgoZimanyi1990,Lisa2005} This method first reconstructs the coordinate-space pair-source distribution $D(\bm{\rho}, \bm{K})$ as an ensemble-accumulated source summed over all events (in contrast to the event-by-event fitting yielding event-averaged parameters in Refs.~\refcite{kincses2025,kincses2026lambda}) and then calculates its Fourier transform via discrete numerical quadrature.

To construct the three-dimensional ensemble-accumulated pair distribution $D(\bm{\rho})$, we select all same-event $\pi^+\pi^+$ pairs from the $0\text{--}10\,\%$ central EPOS3 events that satisfy the single-particle acceptance ($|\eta| < 1.0$, $0.15\text{ GeV}/c < p_{\text{T}} < 1.0\text{ GeV}/c$) within each specified $K_{\text{T}}$ interval.\cite{kincses2025,kincses2026lambda} For each accepted pair, the kinematic variables are transformed into the Longitudinal Comoving System (LCMS), where the average longitudinal momentum vanishes ($K_{z,\text{LCMS}} = 0$). In the Bertsch--Pratt parameterization, the components of the relative momentum $\bm{q} = (q_{\text{out}}, q_{\text{side}}, q_{\text{long}})$ are evaluated as
\begin{align}
q_{\text{out}} &= q_x \cos\phi + q_y \sin\phi\,, \label{eq:q_out} \\
q_{\text{side}} &= -q_x \sin\phi + q_y \cos\phi\,, \label{eq:q_side} \\
q_{\text{long}} &= \frac{2 (p_{1,z} E_2 - p_{2,z} E_1)}{\sqrt{(E_1 + E_2)^2 - (p_{1,z} + p_{2,z})^2}}\,, \label{eq:q_long}
\end{align}
where $\cos\phi = K_x / K_{\text{T}}$ and $\sin\phi = K_y / K_{\text{T}}$. In addition, since the reference data from Refs.~\refcite{kincses2025} and \refcite{kincses2026lambda} were also obtained using the kinematic pair cut $Q_{\text{LCMS}} \equiv \sqrt{q_{\text{out}}^2 + q_{\text{side}}^2 + q_{\text{long}}^2} < \sqrt{150\text{ MeV}/c \cdot m_{\text{T}} c}$, we apply this exact same pair selection to discard wide-angle pairs outside the femtoscopic correlation region and ensure strict consistency and direct comparability.

Because the on-shell condition ($p_1^2 = p_2^2 = m_\pi^2$) implies $q_0 = \beta_{\text{T}} q_{\text{out}} = (K_{\text{T}} / K_0) q_{\text{out}}$ in the LCMS, the four-dimensional quantum-mechanical phase $q \cdot r = q_0 \Delta t - \bm{q} \cdot \bm{r}$ (where $r = x_1 - x_2 = (\Delta t, \bm{r})$ denotes the lab-frame four-separation, labeled as $\rho$ in Eq.~(\ref{eq:D_def})) simplifies to an effective three-dimensional scalar product:
\begin{equation}
q \cdot r = - \left( q_{\text{out}} \rho_{\text{out}} + q_{\text{side}} \rho_{\text{side}} + q_{\text{long}} \rho_{\text{long}} \right) = - \bm{q} \cdot \bm{\rho}\,,
\label{eq:reduced_phase}
\end{equation}
where the three-dimensional reduced relative spacetime coordinates $\bm{\rho} \equiv (\rho_{\text{out}}, \rho_{\text{side}}, \rho_{\text{long}})$ in the LCMS are defined by
\begin{align}
\rho_{\text{out}} &= r_x \cos\phi + r_y \sin\phi - \frac{K_{\text{T}}}{K_0^2 - K_z^2} (K_0 \Delta t - K_z \Delta z)\,, \label{eq:rho_out} \\
\rho_{\text{side}} &= -r_x \sin\phi + r_y \cos\phi\,, \label{eq:rho_side} \\
\rho_{\text{long}} &= \frac{K_0 \Delta z - K_z \Delta t}{\sqrt{K_0^2 - K_z^2}}\,. \label{eq:rho_long}
\end{align}
Since identical pions are indistinguishable quantum-mechanically, the physical pair-source distribution is invariant under spatial inversion, $D(\bm{\rho}) = D(-\bm{\rho})$. To enforce this reflection symmetry statistically in the histogrammed sample, a random sign assignment ($\pm \bm{\rho}$) is applied when filling the distribution. The ensemble-accumulated source $D(\bm{\rho})$ is stored on a cubic grid of $301 \times 301 \times 301$ bins spanning $[-150.5, 150.5]\text{ fm}$ with a uniform cell volume $dV = 1.0\text{ fm}^3$.

After normalizing the source to unity over all bins of the $301^3$ histogram ($\int d^3\bm{\rho} \, D(\bm{\rho}) = 1$), the two-particle correlation function is obtained from the Koonin--Pratt relation:
\begin{equation}
C_2(\bm{q}) = 1 + \int d^3\bm{\rho} \, D(\bm{\rho}) \cos\left( \frac{\bm{q} \cdot \bm{\rho}}{\hbar c} \right)\,,
\label{eq:koonin_pratt_cos}
\end{equation}
where the imaginary sine component vanishes on average due to the statistical reflection symmetry of $D(\bm{\rho})$ and is simply discarded. In the numerical integration code, the integral is computed via discrete Riemann summation over all active spatial cells of the $301^3$ $D(\bm{\rho})$ histogram. To optimize computational efficiency, the summation utilizes OpenMP multithreading alongside SIMD vectorization from the VDT mathematical library.\cite{VDT2014}

For the baseline physics analysis and figures presented in Sec.~3.2, $C_2(\bm{q})$ was evaluated on a $100 \times 100 \times 100$ Cartesian momentum grid spanning $0 < q_i < 150\text{ MeV}/c$ along each axis and fitted over the entire $100^3$ positive-octant cube without any spherical cut using standard least-squares ($\chi^2$) minimization, with bin-by-bin uncertainties propagated from the Poisson errors of the $D(\bm{\rho})$ histogram. In addition, to perform a strict like-for-like comparison with the FFT method, we also performed a dedicated numerical integration run on the exact same momentum grid as that defined by the FFT ($\Delta q_i \approx 4.12\text{ MeV}/c$).

This direct integration approach yields a correlation function free of the pair-sampling and event-grouping fluctuations inherent to CRAB (while retaining the finite-sample statistical fluctuations of the ensemble-accumulated source $D(\bm{\rho})$). However, its computational cost scales as $\mathcal{O}(N_\rho^3 \times N_q^3)$ floating-point operations per $K_{\text{T}}$ bin, requiring approximately four to five hours of execution time per $K_{\text{T}}$ bin.

\subsection{Fast Fourier transformation of the pair-source distribution}

To overcome the heavy computational burden of direct numerical integration, we also evaluate the correlation function using three-dimensional Fast Fourier Transformation (FFT). By mapping the continuous Fourier integral onto a discrete lattice, the computational complexity is reduced from $\mathcal{O}(N_\rho^3 \times N_q^3)$ to $\mathcal{O}(N^3 \log N)$ operations.

We utilize the FFTW3 library\cite{FFTW3} to compute the three-dimensional Discrete Fourier Transform (DFT) on the normalized ensemble-accumulated $D(\bm{\rho})$ distribution defined on the $N^3 = 301^3$ spatial grid ($1.0\text{ fm}$ bin size), where the odd dimension $N = 301$ ensures a symmetric origin bin at $\bm{\rho} = \bm{0}$. By the reciprocity of the DFT, the momentum grid resolution $\Delta q_i$ is fixed by the spatial lattice extent:
\begin{equation}
\Delta q_i = \frac{2\pi \hbar c}{N \Delta \rho_i} \approx \frac{2\pi \times 197.327\text{ MeV fm}}{301 \times 1.0\text{ fm}} \approx 4.12\text{ MeV}/c\,.
\label{eq:reciprocal_grid}
\end{equation}

To re-center $\bm{q} = \bm{0}$ in physical momentum space, we apply a standard index shift ($m = i$ for $i \le (N-1)/2$ and $m = i - N$ for $i > (N-1)/2$). Furthermore, because the FFT assumes the initial array element resides at the origin $\bm{\rho} = \bm{0}$ whereas our input histogram starts at the center of the first bin, $\bm{\rho}_0 = (-150, -150, -150)\text{ fm}$ (with lower bin edge $-150.5\text{ fm}$), the array stores the shifted distribution $D(\bm{\rho} + \bm{\rho}_0)$. Under the $e^{-i \bm{q} \cdot \bm{\rho} / \hbar c}$ convention of FFTW's forward transform (which matches the $e^{+iqx}$ convention of Eq.~(\ref{eq:C2_S}) via $q \cdot r = -\bm{q} \cdot \bm{\rho}$ in Eq.~(\ref{eq:reduced_phase})), the Fourier shift theorem introduces a momentum-dependent complex phase:
\begin{equation}
\mathcal{F}\{D(\bm{\rho} + \bm{\rho}_0)\}(\bm{q}) = \exp\left( i \frac{\bm{q} \cdot \bm{\rho}_0}{\hbar c} \right) \widetilde{D}(\bm{q})\,.
\label{eq:shift_theorem}
\end{equation}
To recover the physical Fourier transform $\widetilde{D}(\bm{q})$, the raw complex output $F(\bm{q})$ is rotated by the phase angle $\phi = \bm{q} \cdot \bm{\rho}_0 / \hbar c$:
\begin{equation}
D_{\text{real}}(\bm{q}) = \mathrm{Re}[F(\bm{q})] \cos\phi + \mathrm{Im}[F(\bm{q})] \sin\phi\,,
\label{eq:phase_rotation}
\end{equation}
where we simply discard the residual imaginary part ($\mathrm{Im}[\widetilde{D}(\bm{q})] \approx 0$) arising from the finite-statistics asymmetry of the random-sign ($\pm\bm{\rho}$) filling and retain only the real part. The correlation function is then obtained as $C(\bm{q}) = 1 + D_{\text{real}}(\bm{q})$ on the FFT momentum grid and, identically to the corresponding numerical integration cross-check run, fitted over the non-negative octant cube $0 \le q_i < 150\text{ MeV}/c$ (including the $q_i = 0$ coordinate planes and the $\bm{q} = \bm{0}$ point, where $C(\bm{0}) = 2$ exactly by normalization) using Eq.~(\ref{eq:C2_Levy_3D}).

\section{Results}

\subsection{Comparison of numerical integration and FFT}

Analytically, the discrete Riemann summation of the Koonin--Pratt integral in Eq.~(\ref{eq:koonin_pratt_cos}) and the phase-corrected DFT in Eqs.~(\ref{eq:shift_theorem}) and (\ref{eq:phase_rotation}) evaluate the exact same discrete Fourier sum over the $301^3$ ensemble-accumulated source $D(\bm{\rho})$. To explicitly validate our numerical implementation and verify the phase-rotation procedure on an equal footing, we performed a direct cross-check between the two methods on an identical momentum grid. For this strict, interpolation-free bin-by-bin comparison of both the correlation functions and the execution times, a dedicated run of the direct numerical integration was executed using the exact same three-dimensional $\bm{q}$-grid ($\Delta q_i \approx 4.12\text{ MeV}/c$) as that generated by the FFT in Eq.~(\ref{eq:reciprocal_grid}).

We evaluated the root-mean-square error (RMSE) between the resulting correlation functions, $C_2^{\text{int}}(\bm{q})$ and $C_2^{\text{FFT}}(\bm{q})$, across all eight $K_{\text{T}}$ intervals and found that the discrepancy remains below $10^{-14}$ in every $K_{\text{T}}$ bin. This residual difference is solely attributable to standard double-precision floating-point rounding errors arising from the different summation orders of the two algorithms. Consequently, three-dimensional L\'evy fits to $C_2^{\text{int}}(\bm{q})$ and $C_2^{\text{FFT}}(\bm{q})$ on the same grid yield identical source parameters ($\lambda, R_{\text{out}}, R_{\text{side}}, R_{\text{long}}$, and $\alpha$), while the FFT reduces the computation time per $K_{\text{T}}$ bin on this identical grid from over four hours to approximately $1.3\text{ s}$.

\subsection{Transverse-mass dependence of the extracted source parameters}

\begin{figure}[t]
  \centering
  \includegraphics[width=0.95\textwidth]{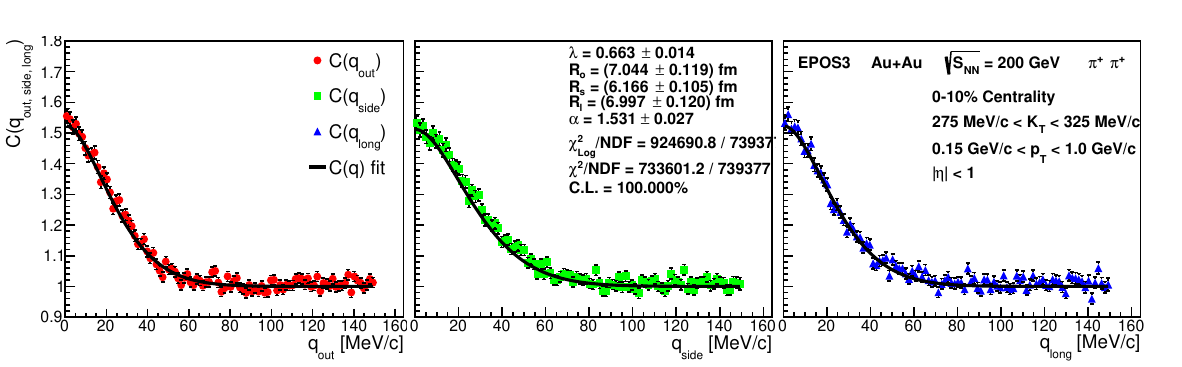}
  \vspace{1ex} 
  \includegraphics[width=0.95\textwidth]{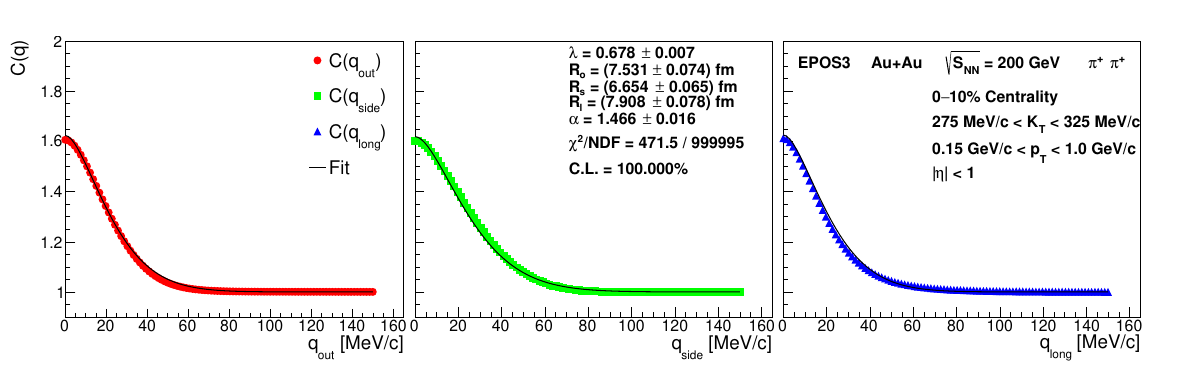}
  \caption{One-dimensional projections of the three-dimensional correlation functions $C_2(\bm{q})$ and their corresponding three-dimensional L\'evy fits (solid curves) along the $q_{\text{out}}$, $q_{\text{side}}$, and $q_{\text{long}}$ axes for a representative $K_{\text{T}}$ interval, obtained using the CRAB package (top panel) and the direct numerical integration/FFT method (bottom panel). For CRAB, a single representative event-grouping realization is displayed, whereas the final parameter values in Figs.~\ref{fig:radii_mT}--\ref{fig:lambda_mT} are obtained by averaging over three different impact-parameter binning configurations.}
  \label{fig:example_fits}
  \alttext{One-dimensional projections of the 3D correlation functions and Levy fits for the CRAB (top) and numerical integration/FFT (bottom) methods.}
\end{figure}

\begin{figure}[t]
\centerline{\includegraphics[width=0.98\textwidth]{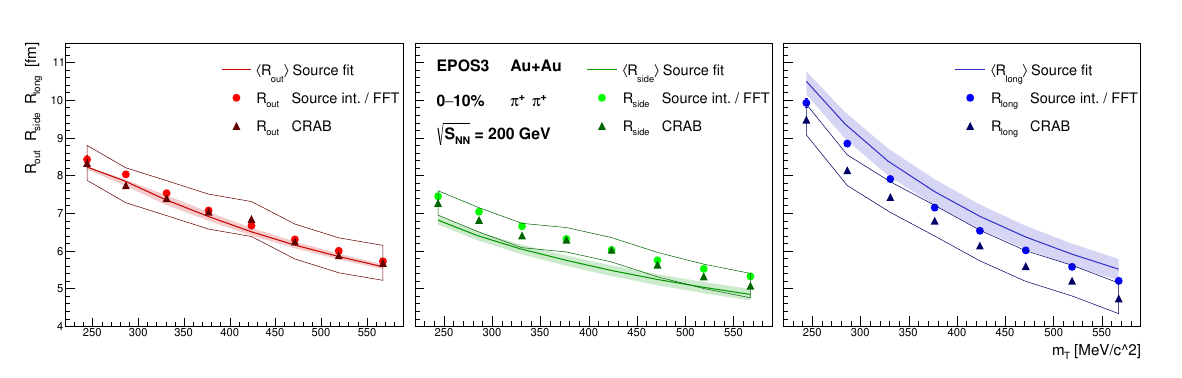}}
\caption{Transverse-mass ($m_{\text{T}}$) dependence of the directional L\'evy scale parameters $R_{\text{out}}$ (left), $R_{\text{side}}$ (middle), and $R_{\text{long}}$ (right) in $0\text{--}10\,\%$ central $\sqrt{s_{\text{NN}}} = 200\text{ GeV}$ $\text{Au}+\text{Au}$ collisions from EPOS3, comparing the event-averaged parameters from direct coordinate-space source fits\cite{kincses2025,kincses2026lambda} (with shaded bands indicating total systematic uncertainties) against CRAB and the numerical integration/FFT method.}
\label{fig:radii_mT}
\alttext{Three-panel plot showing the transverse-mass dependence of R_out, R_side, and R_long across all investigated methods.}
\end{figure}

\begin{figure}[t]
\centerline{\includegraphics[width=0.75\textwidth]{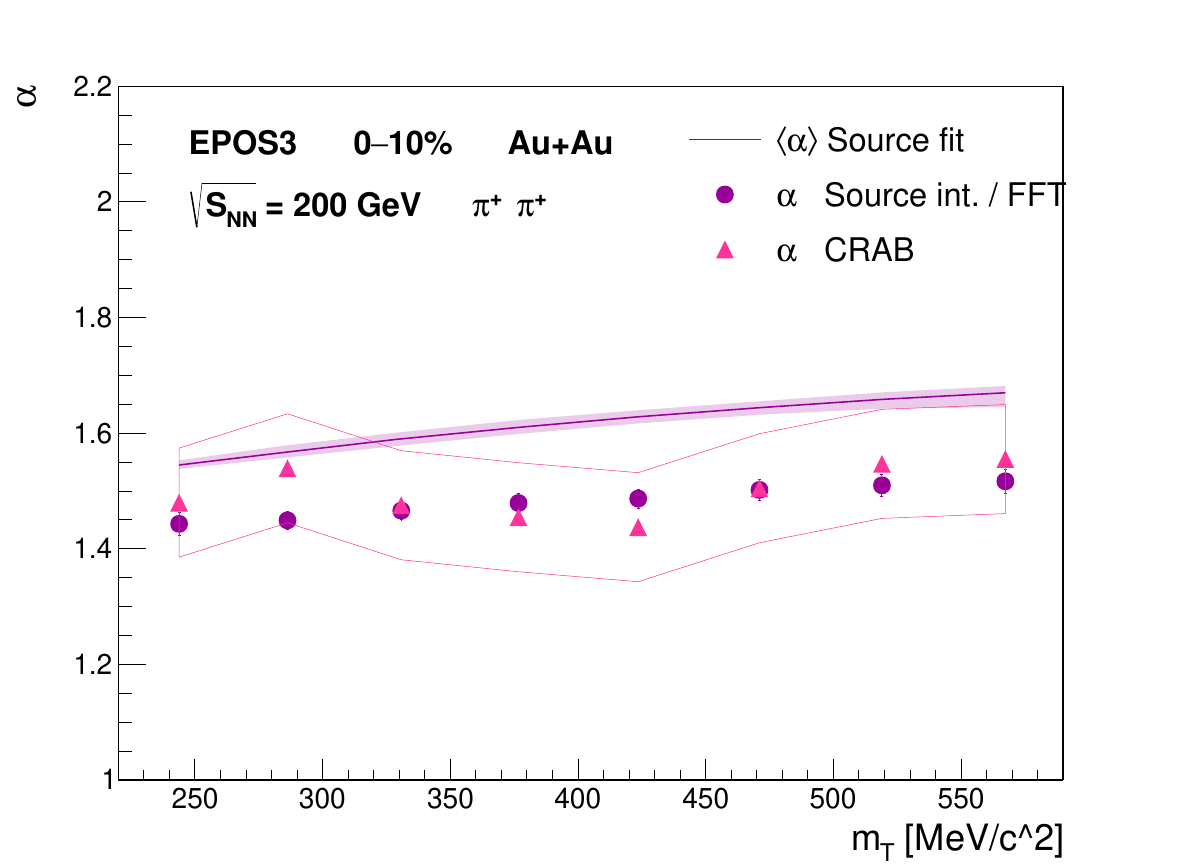}}
\caption{The L\'evy stability index $\alpha$ as a function of $m_{\text{T}}$ extracted via direct pair-source fitting\cite{kincses2025,kincses2026lambda} (shaded bands denoting total systematic uncertainties), CRAB, and the numerical integration/FFT approach.}
\label{fig:alpha_mT}
\alttext{Transverse-mass dependence of the Levy stability index alpha for the compared methods.}
\end{figure}

\begin{figure}[t]
\centerline{\includegraphics[width=0.75\textwidth]{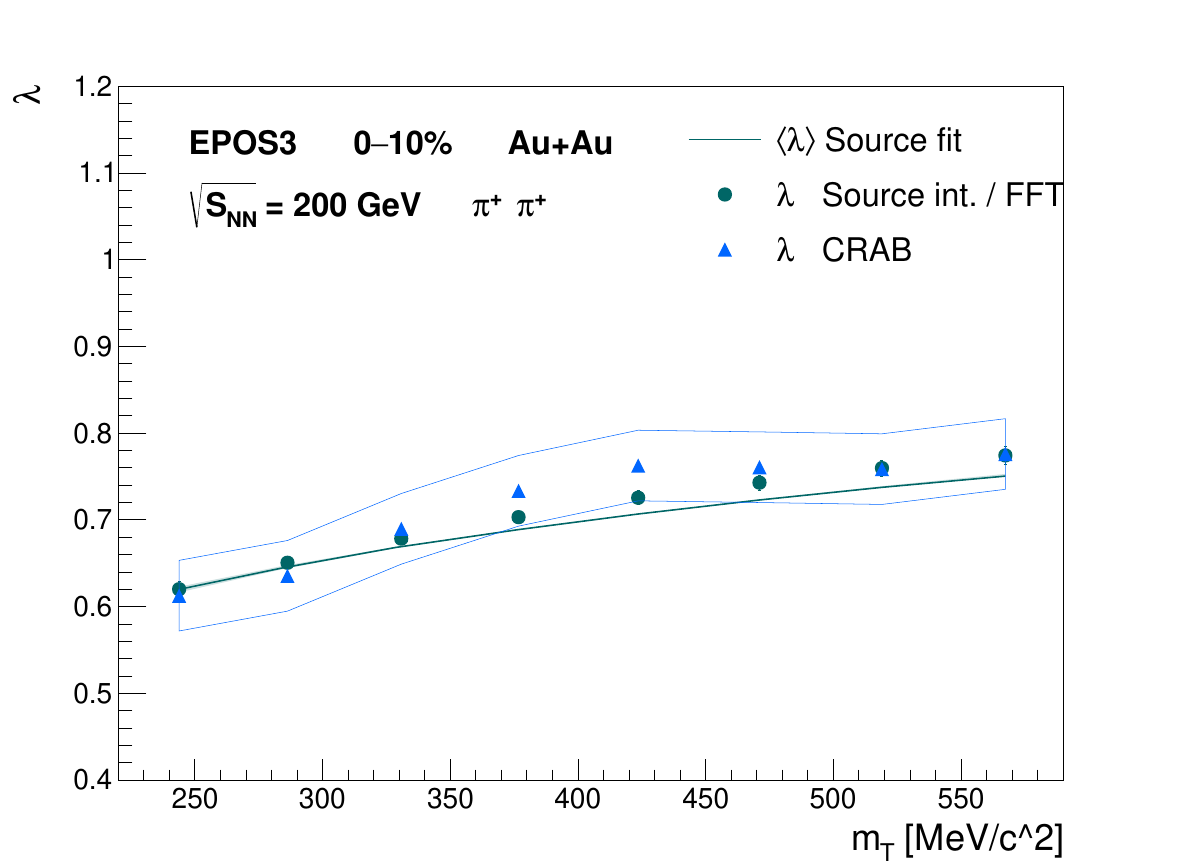}}
\caption{Correlation strength parameter $\lambda$ as a function of $m_{\text{T}}$ compared across the direct source evaluation\cite{Kincses2025shapes,kincses2026lambda} (shaded bands indicating total systematic uncertainties), CRAB, and the numerical integration/FFT method.}
\label{fig:lambda_mT}
\alttext{Transverse-mass dependence of the correlation strength parameter lambda across the compared methods.}
\end{figure}

Following the construction of the three-dimensional correlation functions $C_2(\bm{q})$ in all eight $K_{\text{T}}$ intervals, we extract the physical source parameters ($\lambda$, $R_{\text{out}}$, $R_{\text{side}}$, $R_{\text{long}}$, and $\alpha$) by fitting Eq.~(\ref{eq:C2_Levy_3D}). For the numerical integration results displayed in Figs.~\ref{fig:example_fits}--\ref{fig:lambda_mT}, the correlation functions were evaluated on the $100 \times 100 \times 100$ momentum grid ($0 < q_i < 150\text{ MeV}/c$) and fitted over the full $100^3$ cube with no spherical cut, whereas the CRAB correlation functions (also histogrammed on a $100^3$ grid spanning $0 < q_i < 150\text{ MeV}/c$) were fitted within a spherical domain of radius $|\bm{q}|_{\text{max}} = 200\text{ MeV}/c$. Representative three-dimensional fits for both the CRAB and the numerical integration/FFT methods are illustrated in Fig.~\ref{fig:example_fits} via one-dimensional projections onto the $q_{\text{out}}$, $q_{\text{side}}$, and $q_{\text{long}}$ axes. Since the direct numerical integration and the FFT approach yield identical correlation functions and fit results, they are represented jointly in the subsequent comparisons.

Before comparing the kinematic trends across the different frameworks, it is instructive to examine their respective uncertainties. For the direct coordinate-space fits to $D(\bm{\rho})$ taken from Ref.~\refcite{kincses2026lambda} (continuing the analysis of Ref.~\refcite{kincses2025}), the shaded bands represent the total systematic uncertainty of the event-averaged parameters. In the case of the numerical integration/FFT method, the absence of pair-sampling and event-grouping fluctuations leaves only the finite-sample statistical uncertainties of the ensemble-accumulated source $D(\bm{\rho})$ propagated through the fitting procedure, resulting in very small error bars. However, since all bins of $C_2(\bm{q})$ are computed from the same $D(\bm{\rho})$ histogram, their uncertainties are strongly correlated; consequently, the confidence level shown in Fig.~\ref{fig:example_fits} and the absolute magnitude of the fitted parameter uncertainties are not necessarily statistically interpretable for this method.

By contrast, in the CRAB analysis we observed pronounced point-to-point fluctuations as a function of the transverse mass $m_{\text{T}}$, which stem from the event-grouping and pair-sampling procedure. Specifically, because multiple EPOS3 events within a given impact-parameter ($b$) interval are merged into a single pooled event class from which the numerator pairs in $A(\bm{q})$ are also sampled and assigned Bose--Einstein weights, $A(\bm{q})$ itself includes cross-event pairs and does not sample strictly the same within-event pair source as $D(\bm{\rho})$. Consequently, varying the $b$-bin width significantly alters the extracted parameter values. To account for this sensitivity, we performed the CRAB analysis using three distinct event-grouping configurations---dividing the $0\text{--}10\,\%$ centrality range into 10, 20, and 40 impact-parameter bins. Each final CRAB data point is defined as the mean of the parameter values obtained from these three runs, while its uncertainty is assigned as the $K_{\text{T}}$-bin average of the standard deviations (spreads of the three values) calculated within each individual $K_{\text{T}}$ bin. Averaging the standard deviations across $K_{\text{T}}$ bins is necessary because, with only three runs per bin, the bin-by-bin spread fluctuates strongly depending on how closely the three points happen to cluster in a given bin, which would otherwise lead to unphysically erratic error bars. Despite this averaging, residual point-to-point fluctuations persist across $m_{\text{T}}$; because the fitted parameters are strongly correlated and anti-correlated, these grouping-induced fluctuations appear simultaneously across all extracted parameters.

Figure~\ref{fig:radii_mT} presents the extracted L\'evy radii $R_{\text{out}}$, $R_{\text{side}}$, and $R_{\text{long}}$ as a function of $m_{\text{T}}$. In all three directions, every method reproduces the expected decreasing trend with increasing $m_{\text{T}}$, driven by the hydrodynamic collective expansion of the medium. Comparing the results across the different methods reveals the following direction-dependent behavior:
\begin{itemlist}
\item In the ``out'' direction ($R_{\text{out}}$), all methods show good quantitative agreement across the full $m_{\text{T}}$ range.
\item In the ``side'' direction ($R_{\text{side}}$), a consistent ordering is observed where the direct source fit yields the smallest values, followed by CRAB, while the numerical integration/FFT method gives the largest radii ($R_{\text{side}}^{\text{source fit}} < R_{\text{side}}^{\text{CRAB}} \le R_{\text{side}}^{\text{int/FFT}}$).
\item In the ``long'' direction ($R_{\text{long}}$), the ordering consistently follows $R_{\text{long}}^{\text{CRAB}} < R_{\text{long}}^{\text{int/FFT}} < R_{\text{long}}^{\text{source fit}}$.
\end{itemlist}

The transverse-mass dependence of the L\'evy stability index $\alpha$ is shown in Fig.~\ref{fig:alpha_mT}. The two momentum-space correlation techniques (CRAB and numerical integration/FFT) are mutually consistent, whereas both systematically yield $\sim 7\text{--}9\,\%$ lower $\alpha$ values than the direct coordinate-space source fits ($\alpha^{\text{CRAB}} \approx \alpha^{\text{int/FFT}} < \alpha^{\text{source fit}}$). Furthermore, both the direct source fit and the numerical integration/FFT method exhibit a slight monotonic increase of $\alpha$ with increasing $m_{\text{T}}$, while CRAB follows the same upward tendency within its point-to-point fluctuations. While this behavior contrasts with experimental measurements \cite{PHENIX:2024}, a similar upward trend has been consistently observed in analyses of simulated heavy-ion collision data.\cite{kincses2022,kincses2025,kincses2026lambda}

The behavior of the correlation strength parameter $\lambda$ as a function of $m_{\text{T}}$ is displayed in Fig.~\ref{fig:lambda_mT}. Overall, the numerical integration/FFT results agree with the direct source fit within $\approx 3\,\%$, while the CRAB results exhibit larger, grouping-induced point-to-point fluctuations; toward the highest $m_{\text{T}}$ bins, the numerical integration/FFT method yields slightly higher $\lambda$ values than the direct source fit.

Overall, for interaction-free correlation studies in transport simulations, these comparisons demonstrate that the numerical integration/FFT framework provides a significantly faster and more consistent alternative to CRAB, producing smooth $m_{\text{T}}$ dependencies free of grouping- and pair-sampling-induced fluctuations. Nevertheless, systematic differences remain between the momentum-space fits ($C_2(\bm{q})$) and the direct coordinate-space source fits ($D(\bm{\rho})$), despite using the exact same simulated dataset, single-particle acceptance, and $K_{\text{T}}$ binning. On the one hand, these discrepancies may suggest that the underlying EPOS3 pair source is not strictly of a three-dimensional L\'evy-stable shape, making coordinate-space fits (over $|\rho_i| < \rho_{\text{fit,max}} = \sqrt{B/m_{\text{T}}}$) and momentum-space fits (over $q_i < 150\text{ MeV}/c$) sensitive to different spatial and momentum regimes. On the other hand, the difference (particularly the $\sim 7\text{--}9\,\%$ lower $\alpha$) can also arise from the methodological distinction between the event-averaged parameters ($\langle \alpha \rangle, \langle R \rangle$) of Refs.~\refcite{kincses2025,kincses2026lambda} and fitting the ensemble-accumulated source $D(\bm{\rho})$ summed over all events in the centrality class, since a mixture of L\'evy distributions with event-by-event fluctuating scales and stability indices is not strictly L\'evy and yields a reduced effective $\alpha$ when fitted with a single L\'evy function.

\section{Summary and Conclusions}

In this paper, we performed a methodological comparison for extracting three-dimensional L\'evy femtoscopic source parameters ($R_{\text{out}}$, $R_{\text{side}}$, $R_{\text{long}}$, $\alpha$, and $\lambda$) as a function of transverse mass $m_{\text{T}}$ using $\pi^+\pi^+$ pairs from $0\text{--}10\,\%$ central $\text{Au}+\text{Au}$ collisions at $\sqrt{s_{\text{NN}}} = 200\text{ GeV}$ simulated with the EPOS3 event generator. Using identical kinematic cuts and $K_{\text{T}}$ binning, we compared event-averaged parameters from direct coordinate-space fits to the pair-source distribution $D(\bm{\rho})$\cite{kincses2025,kincses2026lambda} against momentum-space fits to the two-particle correlation function $C_2(\bm{q})$ obtained via: (i) the CRAB afterburner, (ii) direct numerical integration of the Koonin--Pratt formula on the ensemble-accumulated source, and (iii) three-dimensional Fast Fourier Transformation (FFT).

Our main conclusions are as follows:
\begin{romanlist}[(iii)]
\item Evaluating the Koonin--Pratt integral via discrete numerical integration and via phase-corrected 3D FFT on an identical momentum grid from the ensemble-accumulated source yields numerically equivalent correlation functions ($\text{RMSE} < 10^{-14}$ in every $K_{\text{T}}$ bin), while the FFT algorithm reduces the computation time per $K_{\text{T}}$ bin from hours to approximately $1.3\text{ s}$.
\item Compared to CRAB, the numerical integration/FFT approach operates directly on the ensemble-accumulated source and avoids random pair sampling from pooled event classes. In CRAB, merging events into impact-parameter classes introduces noticeable grouping- and pair-sampling-induced point-to-point fluctuations as a function of $m_{\text{T}}$, which propagate into all fitted parameters due to strong parameter correlations and anti-correlations. In contrast, the FFT method produces smooth and consistent $m_{\text{T}}$ trends with minimal statistical uncertainty. At the same time, both methods have distinct scopes and limitations: although each requires the complete freeze-out phase-space coordinates of a model, CRAB mimics the experimental pair-analysis procedure (and can include final-state interactions), whereas the integration and FFT approaches operate directly on the pair-source distribution. Moreover, the $\mathcal{O}(N^3 \log N)$ FFT shortcut relies on the plane-wave kernel depending solely on $\bm{q} \cdot \bm{\rho}$; when Coulomb or strong final-state interactions are included, the kernel no longer reduces to a pure Fourier phase and the FFT shortcut is lost, whereas direct numerical integration remains applicable at the computational cost of several hours per $K_{\text{T}}$ bin.
\item While all methods capture the expected physical $m_{\text{T}}$ dependencies, systematic differences appear between momentum-space correlation fits and direct coordinate-space source fits. Specifically, while $R_{\text{out}}$ and $\lambda$ from the numerical integration/FFT method agree with the direct source fits within $\approx 3\,\%$ (with slightly higher $\lambda$ values at large $m_{\text{T}}$), consistent parameter orderings emerge in $R_{\text{side}}$ ($R_{\text{side}}^{\text{source fit}} < R_{\text{side}}^{\text{CRAB}} \le R_{\text{side}}^{\text{int/FFT}}$) and $R_{\text{long}}$ ($R_{\text{long}}^{\text{CRAB}} < R_{\text{long}}^{\text{int/FFT}} < R_{\text{long}}^{\text{source fit}}$), and momentum-space fits yield L\'evy stability indices $\alpha$ that are systematically $\sim 7\text{--}9\,\%$ lower than those from direct source fits.
\end{romanlist}

Since all methods used the exact same particle sample, binning, and kinematic cuts, the discrepancies between the direct source fits and the integration/FFT correlation fits may stem both from subtle deviations of the EPOS3 pair source from an ideal three-dimensional L\'evy distribution and from comparing event-averaged parameters\cite{kincses2025,kincses2026lambda} with fits to an ensemble-accumulated source (where the superposition of L\'evy distributions with different scales and $\alpha$ values naturally produces a lower effective $\alpha$). Future work will focus on a deeper investigation to disentangle these effects, including a direct comparison between event-by-event and ensemble-accumulated fits as well as the implementation of a genuinely three-dimensional fitting procedure directly on the pair-source distribution $D(\bm{\rho})$.

\section*{Acknowledgments}

This work was supported by NKFIH under grant numbers 2026-4.1.1-MISSZI\'OK-2026-00005, PD-146589, and NKKP ADVANCED 152097.

\bibliographystyle{ws-ijmpa}
\bibliography{bibliography}

\end{document}